\documentclass[12pt]{iopart}
\usepackage{bm,amssymb,graphicx}
\usepackage{iopams}

\begin{document}

\title{Velocity fluctuations of a piston confining a vibrated granular gas}

\author{J. Javier Brey and M.J. Ruiz-Montero}

\address{F\'{\i}sica Te\'orica, Universidad de Sevilla, Apartado
de Correos 1065, E-41080 Sevilla, Spain} \ead{brey@us.es}

\begin{abstract}
The steady state velocity fluctuations of a movable piston located
on the top of a vibrated granular gas are studied by means of
molecular dynamics simulations. From the second moment of the
distribution, a temperature parameter for the piston is defined and
compared with the granular temperature of the gas located just below
it. Then the two temperature parameters refer to two interacting
macroscopic systems. The equipartition of energy valid in usual
molecular systems is strongly violated, and the temperature of the
piston can be larger or smaller than the one of the gas, depending
of the parameters defining the system at the particle level. The
simulation results for the ratio of temperatures are in agreement
with some theoretical predictions from kinetic theory assuming the
validity of a hydrodynamic description in the limit of weak
inelasticity of the gas.
\end{abstract}




\maketitle

One of the characteristic features exhibited by granular fluids is
the violation of the energy equipartition theorem. The granular
temperatures of each of the components of a mixture, defined from
the average kinetic energies of the particles, are not the same.
This was pointed out long ago \cite{JyM87},  and has attracted a lot
of attention in the last decade or so. It is important to stress
that the equality of the temperatures of the components of a mixture
of molecular fluids is not restricted to equilibrium thermal
systems, but also applies to out-of-equilibrium situations, at least
to those that can be described by means of usual hydrodynamics as
derived from kinetic theory or non-equilibrium statistical
mechanics.

For the homogeneous cooling state (HCS) of a binary
granular mixture modeled as an ensemble of smooth inelastic hard
spheres, the ratio of temperatures of the
two components has been obtained from an approximated solution of
the kinetic Enskog equations \cite{GyD99}.  This result is in good agreement with
Molecular Dynamics (MD) simulation data \cite{DHGyD02}. The HCS is
an idealized state in which the system cools homogeneously  and monotonically.
In real experiments, to maintain a granular system
fluidized, external energy must be continuously supplied, and  this is achieved
by means of vibrating or rotating walls, by shearing, by shaking, or by other external fields.
As a consequence, the systems become inhomogeneous.
The non-equipartition has been confirmed by MD simulations of simple shear
flows \cite{GDyH05}, and of vibrated gases \cite{ByT02b,BRyM05}.
There are also some experimental evidences of the coexistence of different
component temperatures in strongly vibrated granular mixtures
\cite{FyM02, WyP02}. Homogeneously driven granular fluids submitted to stochastic forces have also been considered.
The lack of energy equipartition in these systems has been analyzed \cite{ByT02a}, but the
relationship between this kind of ideal driving and actual
experiments is uncertain.

A peculiarity of the above studies is that they are concerned with
quantities going beyond the macroscopic description, in the sense
that the relevant macroscopic local property is the global
temperature (or energy density) of the mixture, and not that of each
of the components. Moreover, the latter are rather difficult to
measure experimentally. Here, a possible experimental set-up  in
which the breakdown of energy equipartition is macroscopically
observed is described and studied by means of MD and kinetic theory.
The experiment is specifically designed to be reproducible in the
laboratory, and might have some practical advantages over previous
related experiments \cite{AMBLyN03}, although it must be stressed that
the study reported here is restricted to low density gases and that no friction
between the particles, of the particles with the piston, or of the latter with the walls of
the container is taken into account.

\begin{figure}
\begin{center}
\includegraphics[scale=0.3,angle=0]{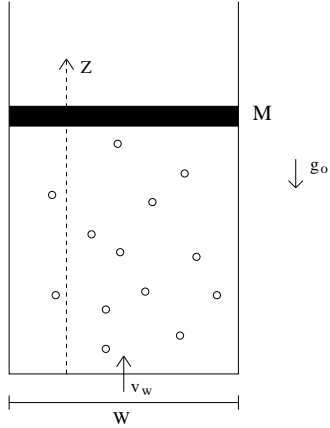} \caption{Sketch of the
system considered in this work.}  \label{fig1}

\end{center}
\end{figure}

Consider a granular system composed of $N$ inelastic hard spheres
($d=3$) or disks ($d=2$) of mass $m$ and diameter $\sigma$. There is
an external gravitational field acting on the system, so that each
particle is submitted to a force $-mg_{0} \widehat{\bm e}_{z}$,
where $g_{0}$ is a positive constant and $\widehat{\bm e}_{z}$ the
unit vector in the direction of the positive $z$ axis. The system is
kept fluidized by vibrating the lower wall of the container in a
sawtooth way, with very high frequency, negligible amplitude, and
velocity ${\bm v}_{W}=v_{W} \widehat{\bm e}_{z}$
\cite{McyL98,BRyM01}. The latter will be always taken large enough
as to keep the density of the system small everywhere. On the top of
the system there is a movable lid or piston of mass $M$, as
illustrated in Fig. \ref{fig1}. The piston can only move in the
$z$-direction, remaining always perpendicular to it. It is assumed
that there is no friction between the piston and the lateral walls
of the vessel containing the granular gas.

Collisions between particles are inelastic and characterized by a
constant coefficient of normal restitution $\alpha$, so that when a
pair of particles $i$ and $j$ collides, their relative velocity ${\bm
v}_{ij}$ changes instantaneously according with
\begin{equation}
\label{1} {\bm v}_{ij} \rightarrow {\bm v}_{ij}^{\prime} = {\bm
v}_{ij} -(1+\alpha) \widehat{\bm \sigma} \cdot {\bm v}_{ij}
\widehat{\bm \sigma},
\end{equation}
while the total momentum is conserved. In the above expression,
$\widehat{\bm \sigma}$ is a unit vector directed from the center of
particle $j$ to the center of particle $i$ at
contact.  Collisions of particles with the piston are also modeled
as smooth and inelastic, being $\alpha_{P}$ the coefficient of
restitution in this case. Taking into account that the piston can
only move in the $z$ direction, the change in the relative velocity
in a collision is restricted to the $z$-component, and it is given by
\begin{equation}
\label{2} v_{z}-V_{z} \rightarrow v_{z}^{\prime}-V_{z}^{\prime} =
v_{z}-V_{z}-(1+\alpha_{P})(v_{z}-V_{z}),
\end{equation}
where $V_{z}$ is the velocity of the piston. This relation and the
total momentum conservation define the collisions with the piston.

Next, some results from MD simulations of two-dimensional systems
will be presented. To avoid undesired effects, periodic boundary
conditions are used in the direction perpendicular to the external
field. It is observed that, at least when the width $W$ of the
system is not too large, a steady state with no macroscopic flow
of mass, and gradients in the fluid only in the direction of the
external field, i.e. the $z$ axis, is reached after some
transitory period. This state is independent from the value of
$W$, as long as it is kept small enough. The shape of the
hydrodynamic profiles in the bulk will be analyzed elsewhere
\cite{ByR08}, while here the emphasis will be put on the
relationship between the temperature parameters of the piston,
$T_{P}$, and of the gas in its vicinity, $T_{G}$. Both
temperatures are defined from the second moment of the respective
velocity distributions, namely  $M \langle V_{z}^{2} \rangle
\equiv T_{P}$, $m  \langle v^{2} \rangle  \equiv 2T_{G}$, with the
angular brackets denoting average. In the results to be reported
in the following, units defined by $m=1$, $\sigma=1$, and
$g_{0}=1$ will be used.

\begin{figure}
\begin{center}
\includegraphics[scale=0.4,angle=-90]{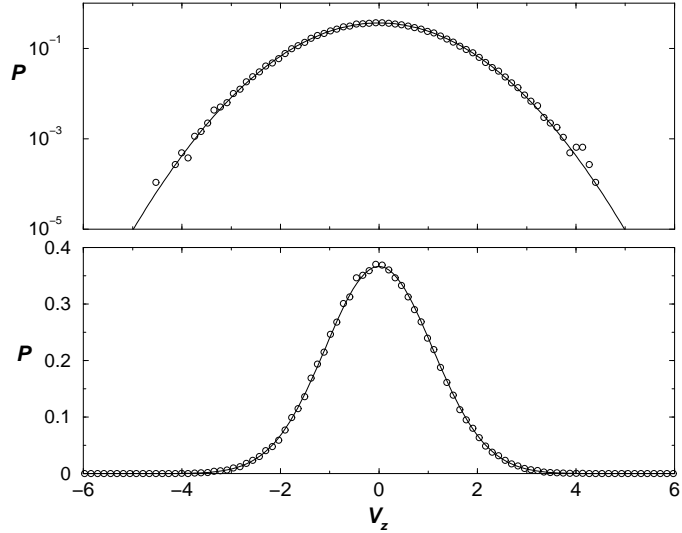} \caption{Steady
velocity distribution of the piston $P(V_{z})$ in both logarithmic
and normal scales. The symbols are from MD simulations while the
solid line is a Gaussian fitting. The values of the parameters
defining the system and the dimensionless unit for the velocity
$V_{z}$ are given in the main text. \label{fig2}}
\end{center}
\end{figure}

The measured probability distribution for  the velocity of the
piston is very well fitted by a Gaussian in all the cases
considered. A typical example, corresponding to $v_{W}=11$, $\alpha=
0.9$, $\alpha_{P}=1$, $M=49$, $N= 420$, and $W=70$, is given in
Fig.\ \ref{fig2}. For this case, the value $T_{P} \simeq 58.67$ is
found. It is seen that the Gaussian fit is accurate at least up to
values of the probability density of the order of $10^{-4}$. The
measurement of $T_{G}$ is somewhat more problematic, since the `top
region' of the granular gas next to the piston has to be identified.
Here, the layer of the gas defined by $\langle Z \rangle - 2
\sigma_{Z} \leq z \leq \langle Z \rangle + 2 \sigma_{Z}$, where
$\sigma_{Z}$ is the dispersion of the values of the position of the
piston $Z$, has been used. Then, the average over the velocities to
compute $T_{G}$ has been carried out over the particles inside this
region at each time. This criterium may appear as rather arbitrary,
but it has been checked that other sensible choices, such as
extrapolating the bulk temperature profile to $\langle Z \rangle$,
lead to values of the temperature with discrepancies of the order of
$ 5 \%$.  In this way, it is found that $T_{G} \simeq 37.69$ ,
implying $T_{P} /T_{G} \simeq 1.56 $. This represents a strong
violation of the energy equipartition, specially taking into account
that the collisions between the grains and the piston have been
modeled as elastic ($\alpha_{P} =1$). Moreover, it must be stressed
that the temperature parameter of the piston is larger than the
temperature of the gas transmitting to it the effects of the lower
vibrating wall. It can be wondered whether this effect is related
with some strong anisotropy of the velocity distribution of the gas.
Let us define $T_{Gx} \equiv m <v_{x}^{2}> $ and $T_{Gz} \equiv m
<v_{z}^{2}>$, so that $T_{G} = (T_{Gx}+T_{Gz})/2$. Then, for the
particular case we are discussing, it is found that $T_{Gz}/T_{Gx}
\simeq 1.096$. Therefore, although there is some anisotropy as usual
in vibrated granular systems, it is very weak as compared with the
excess value of $T_{P}/T_{G}$. More precisely, it is $T_{P}/T_{Gz}
\simeq 1.49$. The same trend is observed in all the other cases
investigated, being the above the one in which the anisotropy of the
second moment of the velocity tensor is stronger.

\begin{figure}
\begin{center}
\includegraphics[scale=0.4,angle=-90]{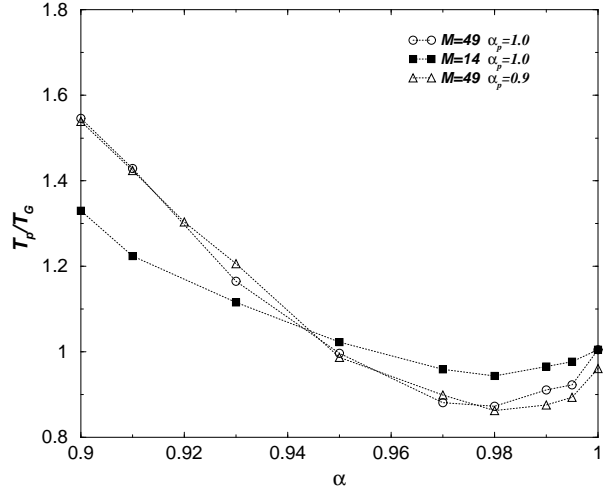} \caption{Ratio of the
temperature of the piston, $T_{P}$, to the temperature of the gas
next to it, $T_{G}$, versus the coefficient of restitution of the
gas, $\alpha$. \label{fig3}}
\end{center}
\end{figure}

In Fig. \ref{fig3}, the temperature ratio $T_{P}/T_{G}$ is plotted
as a function of the coefficient of restitution of the gas $\alpha$
for three different values of the pair ($M$, $\alpha_{P}$),
illustrating the influence of each of these parameters on the
temperature ratio. Quite interestingly, the violation of
equipartition (deviation of the temperature ratio from unity) is not
a monotonic function of $\alpha$, but the temperature ratio exhibits
a minimum in the small inelasticity region. The result with
$\alpha=1$, $\alpha_{P}=1$, has been obtained with a non-vibrated
system at thermal equilibrium. As expected, equipartition is
recovered in the limit $\alpha = \alpha_{P} =1$. The reason to
restrict the analysis to the inelasticity region $\alpha \geq 0.9$
is that, for smaller values, even the hydrodynamic description of
the bulk of the gas, as provided by the Navier-Stokes equations,
breaks down \cite{ByR08}.

An expression for the temperature ratio can be derived as follows.
Given the low density of the gas, a pair of coupled Boltzmann and
Boltzmann-Lorentz kinetic equations are used as the starting point
to describe both the gas and the piston. The subsequent calculations
are based on the following two assumptions: (a) the macroscopic
state of the piston can be specified in terms of the hydrodynamic
fields of the gas in its vicinity and its position probability
distribution, and (b) the associated normal solution of the kinetic
equations can be generated by the Chapman-Enskog procedure. With
these hypothesis, it is obtained that the zeroth order in the
gradients cooling rates of the piston, $\zeta_{P}^{(0)}$ and of the
top region of the gas, $\zeta_{G}^{(0)}$, must be the same
\cite{ByR08,BRyM06}. These rates are functionals of the one-particle
distributions for the top region of the fluid and the piston. Given
that these distributions are not known, Maxwellians can be used as
an estimation. Therefore, no anisotropy effects are introduced in
the theory. A generalization of the derivation in \cite{BRyM06},
with the peculiarity that now all the the collision vectors have the
same direction, leads to \cite{ByR08}
\begin{equation}
\label{3} \zeta_{P}^{*} \equiv \frac{\zeta_{P}^{(0)}}{n_{G} v_{T}
\sigma^{d-1}} = \frac{4  W h}{\pi^{1/2} \sigma^{d-1}}\, (1+
\phi)^{1/2} \left( 1-h \frac{1+ \phi}{\phi} \right),
\end{equation}
\begin{equation}
\label{4} \zeta_{G}^{*} \equiv \frac{\zeta_{G}^{(0)}}{n_{G} v_{T}
\sigma^{d-1}} = \frac{\sqrt{2} \pi^{(d-1)/2}}{ \Gamma \left( d/2
\right) d }\, (1- \alpha^{2}),
\end{equation}
where $n_{G}$ is the gas density in the top region, $v_{T} \equiv (2
T_{G} /m)^{1/2}$ , $\phi \equiv m T_{P}/MT_{G}$, and
\begin{equation}
\label{5} h \equiv  \frac{m(1+\alpha_{P})}{2(m+M)}.
\end{equation}
Note that all the dependence on the mass ratio and the inelasticity
$\alpha_{P}$ occurs through $h$. Requiring the cooling rates to be
the same, an equation determining $\phi$ is obtained,
\begin{equation}
\label{6} h \left( 1 + \phi \right)^{1/2} \left( 1- h
\frac{1+\phi}{\phi} \right)= \frac{(1- \alpha^{2}) \pi^{d/2}
\sigma^{d-1}}{2 \sqrt{2} \Gamma \left( d/2 \right) W d }.
\end{equation}
The elastic collisions limit ($\alpha = \alpha_{P}=1$) is given by
$\phi= h/(1-h) = m/M$, that is the equipartition result. If the
particle collisions are elastic but not those with the piston
($\alpha =1, \alpha_{P} <1$), it is $\phi= m(1+\alpha_{P})/[
2M+m(1-\alpha_{P})]$. This result is the same as the one derived in
ref. \cite{MyP99} for an inelastic intruder in an equilibrium
elastic gas.

\begin{figure}
\begin{center}
\includegraphics[scale=0.4,angle=-90]{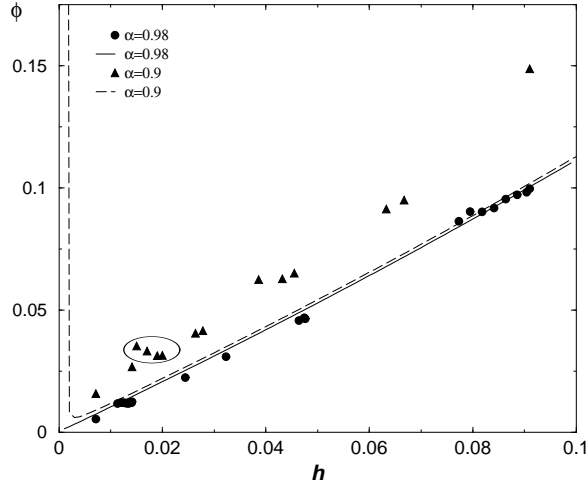} \caption{Ratio $\phi
\equiv mT_{P}/MT_{G}$ of the thermal velocities for the piston and
the gas just below it. The lines are from  Eq.\
(\protect{\ref{6}}) and the symbols MD simulation results for hard
disks. \label{fig4}}
\end{center}
\end{figure}

\begin{figure}
\begin{center}
\includegraphics[scale=0.4,angle=-90]{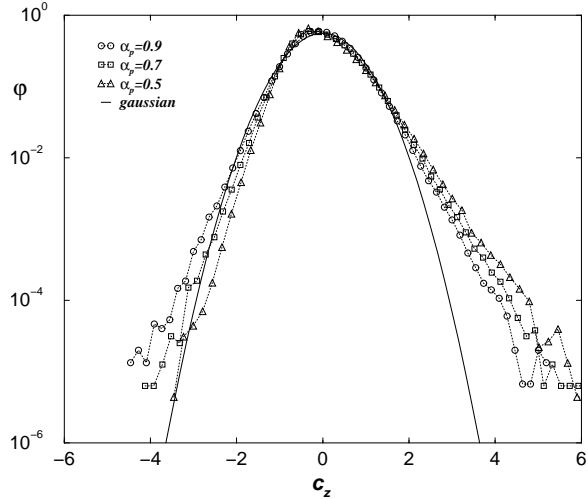} \caption{Marginal
probability distribution of the z-component of the scaled velocity
of the gas, $c_{z} =v_{z}/v_{T}$, in the top region. The values of
all the parameters are the same as in Fig. \protect{\ref{fig2}}.
\label{fig5}}
\end{center}
\end{figure}

The prediction made by Eq. (\ref{6}) is compared with some MD
simulation results in Fig. \ref{fig4}, where $\phi$ is plotted as a
function of $h$. In the simulations, the values of the latter
quantity have been modified by changing both $M$ (in the range
$10$-$140$) and $\alpha_{P}$ (between $0.6$ and $1$). The other parameters
($W$, $v_{W}$, and $N$) are kept the same as in Fig. \protect{\ref{fig2}}.
The symbols
are the simulation results while the lines are from Eq. (\ref{6}),
as indicated. Data for two values of $\alpha$, $0.98$ and $0.9$, are
reported. For the most elastic gas, the agreement is quite good, for
all the considered values of the other parameters. Nevertheless,
when the inelasticity of the gas is slightly increased, strong
deviations from the theoretical prediction show up. These
discrepancies are not only quantitative, but also qualitative. The
circled symbols on the left lower corner of the figure correspond
all them to $M=49$, differing in the value of $\alpha_{P}$. It
is observed that when $\alpha_{P}$ (and $h$) increases, $\phi$
decreases, in contradiction with  the solution of Eq. (\ref{6}). To
identify the cause of the failure of the theory, the shape of the
marginal distribution of the component of the velocity of the grains
perpendicular to the piston in the top region, $\varphi(v_{z})$, has been
measured. While for $\alpha = 0.98$ it is
always well fitted by a Gaussian within the numerical error of the simulations, for
$\alpha = 0.9$, exponential tails are clearly identified for positive
velocities already in the thermal range, as it is seen
in Fig.\ \ref{fig5}. These tails are relevant for the lowest
velocity moments of the gas. On the other hand, it was already mentioned above
(see Fig. \protect{\ref{fig2}}) that the
velocity distribution of the piston remains Gaussian. A theoretical analysis of the existence of this
exponential tail probably requires going beyond the Chapman-Enskog method,
the restriction to normal solutions of the
Boltzmann equation and, therefore, of the hydrodynamic description.

The above results seem to provide additional evidence of the
difficulties when trying to give a macroscopic meaning to the
granular temperature, at least as defined in kinetic theory and the
usual hydrodynamic theories. The knowledge of the hydrodynamic
profiles of the confined gas and the characteristics of the
interactions between the gas particles and the piston is not enough
to determine the ``temperature'' of the latter. More detailed
information about the velocity distribution of the gas in its
vicinity is required. Moreover, it is surprising that sometimes the
``temperature'' of the body to which energy is supplied (the piston)
be larger than the temperature of the system supplying the energy
(the neighbor gas). The results also show that this effect is not
associated to the anisotropy of the gas. Finally, let us stress the
differences of this system as compared with a gas mixture, even in
the tracer (one-intruder) limit. Here, two macroscopic systems
spatially differentiated are interacting at contact, their
respective states being strongly interdependent. As a consequence,
it seems clear that both temperature parameters have a macroscopic
meaning.

\ack

This research was supported by the Ministerio de Educaci\'{o}n y
Ciencia (Spain) through Grant No. FIS2005-01398 (partially
financed by FEDER funds).

\end{document}